%% file: user_study_paper.tex
\documentclass[sigconf]{acmart} 
\newif\ifanonymous
\anonymousfalse 
\usepackage[utf8]{inputenc} 
\usepackage{booktabs} 
\usepackage{array}    
\usepackage{tabularx} 
\usepackage{hyperref}
\usepackage[nameinlink,capitalise,noabbrev]{cleveref}
\crefname{lstlisting}{Listing}{Listings} 
\Crefname{lstlisting}{Listing}{Listings}
\usepackage{multirow}
\usepackage[inline]{enumitem} 
\newlist{inlist}{enumerate*}{1}
\setlist[inlist]{itemjoin={{, }},itemjoin*={{, and }},label=($\roman*$),mode=boxed}
\usepackage{csquotes} 
\let\say\enquote
\usepackage{textcomp} 
\usepackage{listings}
\usepackage{fancyhdr} 
\usepackage{xcolor}
\usepackage{caption}
\usepackage[detect-all]{siunitx}
\PassOptionsToPackage{protrusion=true, activate={true,nocompatibility}, final, tracking=true, kerning=true, spacing=true, factor=1100}{microtype}
\usepackage{etoolbox}  
\usepackage{microtype}
\usepackage{xspace} 
\newcommand{\dash}{\thinspace---\thinspace\xspace}
\usepackage{orcidlink} 
\usepackage[subtle]{savetrees} 
\def\hyph{-\penalty0\hskip0pt\relax}

\DeclareRobustCommand\HypothesisLabel[1]{\ifcase#1
\or $H_{01}$
\or $H_{A1}$
\or $H_{02}$
\or $H_{A2}$
\else ???\fi}
\newcommand{\RsqTable}{\rule{0pt}{2.5ex}$R^2$} 
\newcommand{\Rsq}{$R^2$}

\newcommand{\var}[1]{\textit{#1}}

\long\def\AnswerBox#1#2{\smallskip\par\noindent\fbox{\parbox{\dimexpr\linewidth-2\fboxsep-2\fboxrule}{\textbf{Answer to \ref{rq:#1}}:~\textit{#2}}}\medskip\par}

\acmSubmissionID{149}
\acmConference[EASE 2025]{The 29th International Conference on Evaluation and Assessment in Software Engineering}{17–20 June, 2025}{Istanbul, Türkiye}

\fancypagestyle{firstpage}{
  \fancyhf{}
  \lhead{This is the author’s version of the manuscript accepted for publication at EASE 2025. The final version is available at ACM.}
  \rhead{}
}

\begin{document}
\newcommand{\savedlhead}{\leftmark}
\newcommand{\savedrhead}{\rightmark}

\title{Providing Information About Implemented Algorithms Improves Program Comprehension: A Controlled Experiment}

\author{Denis Neumüller\,\orcidlink{0000-0003-3872-0188}} 
\affiliation{
  \institution{Ulm University}
  \country{Germany}
}
\email{denis.neumueller@uni-ulm.de}

\author{Alexander Raschke\,\orcidlink{0000-0002-6088-8393}}
\affiliation{
  \institution{Ulm University}
  \country{Germany}
}
\email{alexander.raschke@uni-ulm.de}

\author{Matthias Tichy\,\orcidlink{0000-0002-9067-3748}}
\affiliation{
  \institution{Ulm University}
  \country{Germany}
}
\email{matthias.tichy@uni-ulm.de}

\begin{abstract}
\textbf{Context:}
Various approaches aim to support program comprehension by automatically detecting algorithms in source code.
However, no empirical evaluations of their helpfulness have been performed.
\textbf{Objective:}
To empirically evaluate how algorithm labels \dash which include the algorithm's name and additional information \dash impact program comprehension in terms of correctness and time.
\textbf{Method:} 
We conducted a controlled experiment with 56 participants, where the experimental group received code with labeled algorithms.
The groups completed exercises designed to measure program comprehension as well as a post-questionnaire on label helpfulness, use cases for algorithm recognition, and reasons for self-implementation of algorithms in practice.
\textbf{Results:} 
Annotating source code with algorithm labels significantly improves \var{program comprehension} (\textit{p}=0.040), with a median improvement of 6 points (\textasciitilde23\%),
but does not affect completion \var{times} (\textit{p}=0.991).
Qualitative analysis revealed that a majority of participants perceived the labels as helpful, especially for recognizing the codes intent.
Participants also proposed use cases such as error detection, optimization, and library replacement.
Reasons for self-implementing algorithms included library inadequacies, performance needs and avoiding dependencies or licensing costs.
\textbf{Conclusion:}
This study shows that algorithm labels improve program comprehension, especially for developers with medium programming experience. 
Our qualitative analysis also sheds light on how participants benefit from the labels, further use cases for algorithm recognition and motivations behind self-implementing algorithms.
\end{abstract}



\maketitle
\thispagestyle{firstpage}
% After the first page, restore the original header content
\fancypagestyle{plain}{
  \fancyhf{}
  \lhead{\savedlhead}  % Restore the saved left header
  \rhead{\savedrhead}  % Restore the saved right header
}

\section{Introduction}\label{sec:Introduction} 
Program comprehension plays a crucial role in software development, with empirical studies showing that developers spend over half of their time understanding code~\cite{Xia2018, Minelli2015}. 
A variety of approaches, usually referred to as Cliché-, Plan-, Concept- or Algorithm-recogni\hyph tion,
aim to support program comprehension by automatically recognizing language-independent ideas of computation and problem-solving methods, 
such as algorithms and coding strategies used in source code~\cite{Kozaczynski1992, Quilici1994, Wills1994, Metzger2000, Alias2003, Zhu2011, Taherkhani2013, Mesnard2016, Nunez2017, ARCC2, Long2022, Neumueller2024}. 
However, to the best of our knowledge, none of these approaches actually perform an evaluation of their impact on program comprehension.
The few evaluations that exist only consider technical aspects, such as scalability.

This work's goal is to evaluate the effectiveness of these approaches in supporting program comprehension.
Given that algorithms represent some of the more intricate patterns detected by these tools, our study specifically focuses on them. 
We follow Cormen et al.~\cite[p. 5]{DBLP:books/daglib/0023376} and define algorithm as \textit{\say{a specific computational procedure~[\ldots] for solving a well-specified computational problem}}, e.g., sorting, handling data structures, computing shortest paths in a graph, etc. 
While our study specifically focuses on algorithms, we believe the findings also provide valuable evidence for the effectiveness of recognizing computational patterns and concepts in source code, since algorithms can be seen as special types of these.
To ensure a tool-agnostic evaluation, we present algorithm information in a generic way by inserting comments into the code base.
These comments, referred to as algorithm labels in the following, consist of the algorithm's name as well as a Wikipedia link for further reference.
An example of our algorithm labels can be seen in~\cref{lst:levenshtein}. 
The goal of our study can be summarized using the GQM template~\cite{Jedlitschka2005} as follows:
\textbf{Analyze} algorithm labels
\textbf{for the purpose of} evaluation
\textbf{with respect to} their influence on program comprehension
\textbf{from the point of view of} researchers and programmers
\textbf{in the context of} students, researchers and software developers using Java.

From this goal we derive the following research questions:
\begin{enumerate}[label={\textbf{RQ\arabic*:}},ref={RQ\arabic*},leftmargin=*]
  \item \label{rq:1}Does annotating the source code with labels of the implemented algorithms improve program comprehension?
  \item \label{rq:2}Does annotating the source code with labels of the implemented algorithms improve the time required to comprehend a program? 
  \item \label{rq:3}What qualitative insights can we gain from the participants regarding the influence of algorithm labels.
\end{enumerate}
To answer our RQs we conducted a controlled experiment with 56 participants, including students, researchers and industry developers.
Participants were assigned to either the control group or the experimental group, which received algorithm labels as comprehension support.
The groups completed exercises in a realistic Java code base designed to measure program comprehension allowing us to answer \ref{rq:1} and \ref{rq:2}.
For \ref{rq:3}, we predefined 3 analysis dimensions exploring them through the post-questionnaires and interviews:
\begin{inlist}
	\item the perceived usefulness of labels
	\item use cases for algorithm recognition
	\item reasons for self-implementation of algorithms
\end{inlist}.

The contributions of our paper include an evaluation of the impact of algorithm labels on program comprehension in terms of correctness and time,
as well as qualitative insights into why these labels are helpful, use cases for algorithm recognition, and reasons for self-implementation of algorithms in practice.

\section{Experimental Design}
We present the design and findings of our study following the guidelines on reporting empirical studies in Software Engineering outlined by Jedlitschka et al.~\cite{Jedlitschka2005}.
The design and setup of our study follows the recommendations by Siegmund~\cite{Siegmund2012a} and Dunsmore et al.~\cite{Dunsmore2000}, who outline best practices for designing program comprehension experiments, such as controlling confounding variables.
A replication package for our work is available on Zenodo~\cite{neumueller2025Zenodo}.

\subsection{Goals}
\begin{figure*}[t]
	\centering
	\includegraphics[width=0.8\textwidth, trim=0 0 0 0, clip]{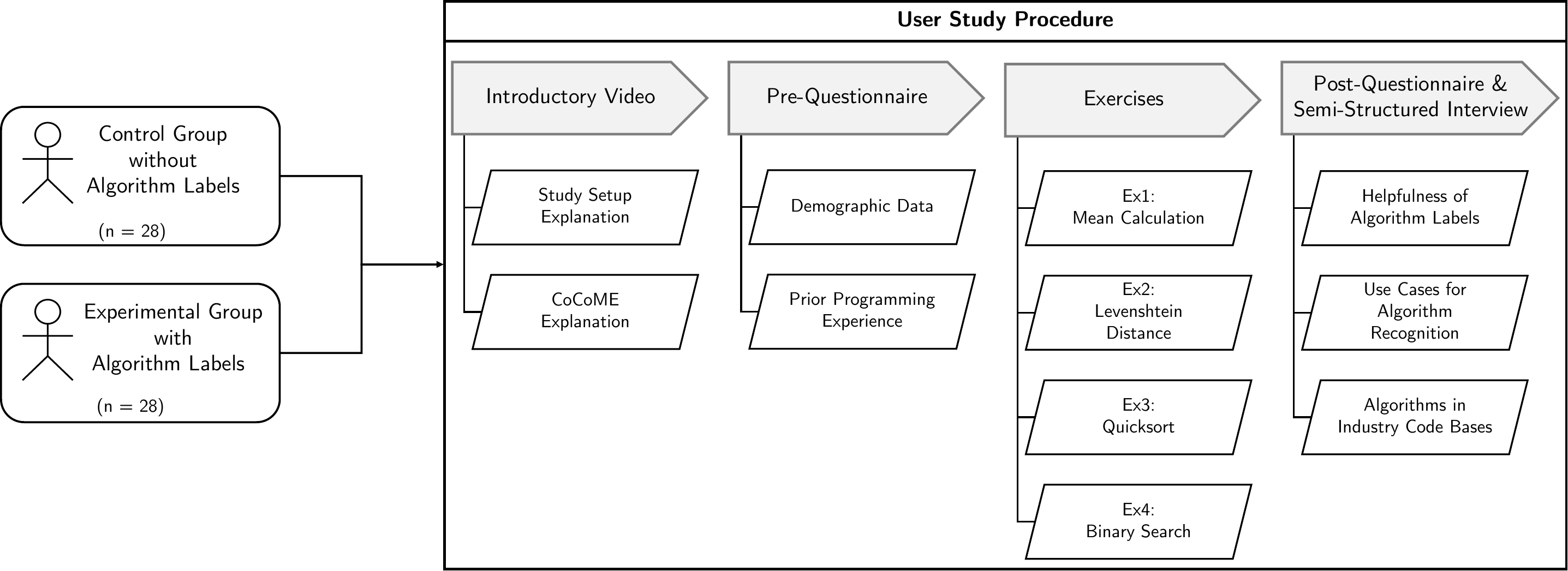}
	\caption{Overview of the study procedure.}
	\label{fig:Overview}
\end{figure*} 
\Cref{fig:Overview} gives an overview of the design of our study.
As described in \cref{sec:Introduction} the goal of our experiment was to find out if annotating source code with labels regarding the implemented algorithms improves program comprehension or the time required to comprehend programs.
To investigate this we designed a controlled experiment 
dividing our participants into a control and experimental group.

Initially, participants watched an introductory video that outlined the study procedure, including the types of exercises and the think-aloud methodology.
It also introduced the CoCoME codebase (see Sec. \ref{sec:Object}), providing an overview of its structure and demonstrating how to launch the software and run its unit tests in Eclipse.

We then collected demographic and educational characteristics about the participants based on a questionnaire designed by Siegmund~\cite[p. 83]{Siegmund2012a} which we adopted almost one-to-one.
Since our study also included participants from industry we added questions on job title and working experience. 
To assess participants' prior programming experience, we utilized a set of programming questions modeled after those developed by Siegmund~\cite[p. 193-199]{Siegmund2012a}.

Thereafter, the actual experiment began in which the participants had to complete four exercises consisting of program comprehension and maintenance exercises.
During these exercises the source code of the experimental group was annotated with the algorithm labels such as the one displayed in~\cref{lst:levenshtein}.

After the experiment the participants had to answer a post-question\hyph naire designed to investigate whether and how the labels helped the subjects and what subjects thought about the algorithm labels.

\subsection{Hypotheses} \label{sec:Hypotheses}
Regarding \ref{rq:1} we expected that providing participants with algorithm labels reduces the need to recognize or infer the implemented algorithms themselves, minimizing confusion and errors. 
Thereby facilitating the understanding of the code's functionality and improving the accuracy of answers to related questions.

With respect to \ref{rq:2} our expectation was that the labels would enable participants to use a top-down comprehension strategy instead of performing time-consuming bottom-up comprehension. 
Finally, we also expected the labels to enable participants to quickly look up information about the algorithm if necessary, improving both comprehension and potentially saving time on exercises.
We thus formulated the following hypotheses regarding our research questions:

\begin{enumerate}[label={\HypothesisLabel{\arabic{enumi}}:}, ref={\HypothesisLabel{\arabic{enumi}}}, leftmargin=4em]
    \item \label{H01} The correctness of the answers to the given exercises \textit{does not significantly differ} if the source code is annotated with labels of the implemented algorithms.
    \item \label{HA1} The correctness of the answers to the given exercises \textit{does significantly differ} if the source code is annotated with labels of the implemented algorithms.
    \vspace{1mm}
    \item \label{H02} The time required to complete the given exercises \textit{does not significantly differ} if the source code is annotated with labels of the implemented algorithms.
    \item \label{HA2} The time required to complete the given exercises \textit{does significantly differ} if the source code is annotated with labels of the implemented algorithms.
\end{enumerate}

Note that our hypotheses are formulated in a two-sided way, which means that the alternative hypothesis is also accepted if the experiment group actually performs worse than the control group.
This allows us to detect the \dash in our opinion unlikely \dash case that the provided algorithm labels have a negative effect.

\subsection{Parameters and Variables} 
\Cref{tab:variables} presents the relevant variables in our experiment, including scales, units, and value ranges.
\var{Group} affiliation, serving as the independent variable, was determined prior to the experiment.
Participants were assigned either to the control or experimental group, balancing group sizes and the distribution of students and employees.

The dependent variables where measured once per exercise and aggregated into a total.
The control variable was measured in the pre-questionnaire and used as a covariate in our statistical analysis,
since we expect that the pre-existing programming experience of participants will have a significant effect on their performance.

We operationalized the measurement of \var{program comprehension} and prior \var{programming experience} using graded exercises based on the guidelines of Siegmund~\cite{Siegmund2012a} and Dunsmore et al.~\cite{Dunsmore2000}.
To ensure a reliable evaluation, two researchers independently graded the exercises and awarded points using pre-established sample solutions and written assessment guidelines, resolving discrepancies through discussion.
We also quantified our inter-rater agreement using an analysis based on Altman et al.~\cite{Bland1999},
aggregating and averaging the absolute rater differences across the questions used to measure \var{program comprehension} and prior \var{programming experience}. 
No rating differences occurred for the latter.
For \var{program comprehension}, the average rater difference was 0.40 points (95\% percentile range [0, 1.94]). 
Since these differences are minor compared to the mean score of 27.95, we consider this to be very good agreement.
\begin{table}[h]
  \centering
  \renewcommand{\arraystretch}{1.25} 
  \begin{tabular}{@{}>{\raggedright\arraybackslash}l@{\hskip8pt}l*3{@{\hskip5pt}l}@{}}
  \toprule
  \textbf{Type} & \textbf{Variable} & \textbf{Scale} & \textbf{Unit} & \textbf{Range} \\ 
  \midrule 
  Independent & Group & Nominal & N/A & \parbox{1.8cm}{\strut Control,\\Experimental} \\ \hline
  \multirow{2}{*}{Dependent}& \parbox{1.9cm}{\strut Program com-\\prehension} & Ratio & Points & [0, 40.5] \vspace{2mm}\\
  & Time & Ratio & Minutes & [0, 98] \\ \hline
  Control & \parbox{1.9cm}{\strut Programming\\experience} & Ratio & Points & [0, 6.5]\\ \bottomrule
  \end{tabular}
  \caption{Description of Variables.}
  \label{tab:variables}
\end{table}

\subsection{Subjects}
To find a sufficiently large and diverse set of participants, we approached students and researchers at the computer science faculty of our as well as other universities, and our industrial contacts.
Our participants can be split into three demographics, consisting of students, academic- and industry employees.
To improve external validity we required that our participants were programming regularly and, for students, to be at least in the fifth-term of their bachelor's program.

In total, 56 participants took part in our study: 25 students, 16 academics, and 15 industry employees.
The students, all studying computer science or related fields like software engineering, enrolled in a median of ten programming-related courses.
The academics consisting primarily of PhD students and three postdocs, reported a median work experience of 5 years. 
The industry participants i.e. professional developers had a median work experience of 6 years.
Across all demographic groups 20 participants had experience with larger programming projects involving 15 or more team members, with a median duration of 3 years.
Of the participants, 36 attended in person, while 20 participated remotely, with a higher proportion of industry employees in the remote group.

To motivate and thank the participants, each received a bar of chocolate. Additionally, they could choose between €20 or 2 participant hours.
Participant hours were only relevant for students enrolled in the \say{Empirical Research Methods} course, as they counted toward fulfilling the required study participation time. 
Students were free to choose which studies to take part in to meet this requirement.

\subsection{Object}\label{sec:Object}
As a source code base we used CoCoME which stands for \say{Common Component Modeling Example}.
CoCoME was developed as part of the DFG Project \textit{Design For Future - Managed Software Evolution} with the goal to serve as a realistic example of a company application for the research project~\cite{Reussner2019}.
More precisely, CoCoME represents the distributed trading system of a supermarket chain, handling use cases like processing customer sales, resupplying individual stores from the central warehouse and generating reports.

To design our exercises we relied on the study of Dunsmore et al.~\cite{Dunsmore2000} which investigated different program comprehension measures, concluding that mental simulation and maintenance exercises were the most effective.
We therefore based our exercises on these types of measures, designing two exercises in which the participants had to answer questions about the code that required mental simulation.
We also designed two maintenance exercises in which the participants had to debug and fix a fault in the software as well as shortly describe the problem, and its solution. 
While the code for exercise one was already part of the system the other exercises were carefully integrated into the system in the context of realistic use cases.
Aside from being good to measure program comprehension we also believe that debugging and reasoning about code execution are common programming activities, strengthening the external validity of our study.

The first of our \textbf{mental simulation exercises} is based on the use case of a Delivery Report which was already implemented in CoCoME. 
It calculates the mean time to delivery for each supplier by querying the order- and delivery-dates for each product, subtracting and aggregating these values and dividing by the number of orders. 
We chose this use case since the first exercise also served as an onboarding exercise and calculating the mean is a simple algorithm.

The second mental simulation exercise is based on the use case of a product search. 
\Cref{lst:levenshtein} presents one of the methods included in this functionality, along with the algorithm label whose influence we examine. 
Both mental simulation exercises required participants to answer program comprehension questions about the meaning and purpose of variables and the output of the programs given certain inputs.
\begin{figure}[h] 
\centering
\begin{lstlisting}[caption={The method implementing the Levenshtein Distance algorithm from exercise two. Note that the comment above the method is the algorithm label that was only provided to the experimental group.},
  label={lst:levenshtein}, captionpos=b, xleftmargin=10pt, numbersep=5pt, linewidth=\dimexpr\linewidth+5pt\relax,]
/* Algorithm: Levenshtein Distance
   https://en.wikipedia.org/wiki/Levenshtein_distance
*/
public static int getDist(CharSequence lhs, CharSequence rhs) {                          
    int len0 = lhs.length() + 1;                                                     
    int len1 = rhs.length() + 1;                                                     
    
    int[] cost = new int[len0];                                                     
    int[] newcost = new int[len0];                                                  
    
    for (int i = 0; i < len0; i++) {
        cost[i] = i;                                     
    }
    
    for (int j = 1; j < len1; j++) {                                                
        newcost[0] = j;                                                             
        
        for (int i = 1; i < len0; i++) {                                             
            int match = (lhs.charAt(i - 1) == rhs.charAt(j - 1)) ? 0 : 1;
            int cost_replace = cost[i - 1] + match;                                 
            int cost_insert  = cost[i] + 1;                                         
            int cost_delete  = newcost[i - 1] + 1;

            newcost[i] = Math.min(Math.min(cost_insert, cost_delete), cost_replace);
        }                                                                                                                          
        int[] swap = cost; 
        cost = newcost; 
        newcost = swap;                          
    }                                                                                  
    return cost[len0 - 1];                                                          
  }
\end{lstlisting}
\end{figure}

The first of the \textbf{maintenance exercises} is based on the use case of a product margin report, that can be used to assess and compare the profitability of different products.
It lists all products with related details such as the supplier and calculates each product's profit margin.
The Quicksort algorithm is used to sort the products in the report based on their margin.
The implementation included an error that excluded the pivot element in subsequent iterations, leaving it unsorted.
Depending on the initial order of the input the implementation would sometimes appear to sort correctly while failing for other inputs.

The second maintenance exercise is based on the use case of querying a specific product contained in an order using its ID.
Since the different products in an order are already sorted by their ID the existing functionality uses the binary search algorithm to find the product.
The implementation included an off-by-one error that lead to the exclusion of the highest element for each sub-list.
This lead to a failure if the excluded element was the one being searched for.

For both maintenance exercises we provided the participants with a test suite that contained both passing and failing tests.
We also carefully adjusted the available time and achievable points for each exercise based on its complexity to ensure comparability across exercises.

\subsection{Instrumentation}
Taking part in the study was possible in two ways: either in person or remotely.
The former was chosen by all faculty staff and almost all the students.
In this case the study took place in a separate office room, which provided a quiet environment to minimize distractions.

In terms of technical setup, we selected Eclipse as the IDE since it is well-known, widely used and open-source.
Indeed, most (n=45) of our participants reported that they \textit{\say{had used it a few times}} in the past.
We used Limesurvey to collect participants' answers and automatically track the time required to complete the exercises. 
Additionally, we recorded a screen capture of the actions performed by the participants as well as their think aloud vocalization for qualitative analysis.

To increase participation, especially from industry, we also provided the possibility of taking part remotely.
For this we used the NoVNC tool, enabling participants to connect remotely to the office machine through their browser and use the same software setup as local participants.
To communicate with these participants video conference software e.g. Zoom was used. 
While we could not control the environment and hardware setup for remote participants, we argue that they took part from their home or work office, using their familiar hardware setup thereby strengthening external validity.

Participants were explicitly instructed to approach the exercises as they would in a real-world setting, using the techniques they typically use for debugging or code comprehension.
They were also allowed to use external resources, such as searching the web, reflecting how developers typically address issues in practice.

\subsection{Data Collection Procedure}
The experimental execution was conducted between February and June of 2024 in one-to-one settings.
During the study, the researchers' role was primarily to remind participants when they were approaching the end of the allotted time for an exercise.
If participants took too long to complete an exercise, they were instructed to move on to the next one.
Additionally, the researchers took notes on interesting observations throughout the experiment such as participant behavior and how participants interacted with the algorithm labels.

The user study was designed to last approximately two hours in order to minimize participant fatigue and concluded with a semi-structured interview. 
This allowed the researchers to ask follow-up questions about interesting aspects of the study process and the participants' responses to the post-questionnaire.

\subsection{Analysis Procedure}
Because we were interested in determining if our treatment has a significant effect while controlling for programming experience we initially planned to perform a Multiple Analysis of Covariance (MANCOVA)~\cite[p. 63]{dattalo2013analysis}.
However since the assumption of normality was violated we instead used PERMANCOVA~\cite{Anderson2017}, which is a robust non-parametric alternative widely used in fields like ecology~\cite{Anderson2017}~\cite[p. 158]{Bakker2024}.

\subsection{Deviations} 
Unforeseen deviations could not be avoided completely, with five participants experiencing a browser crash.
While we promptly resumed each questionnaire, LimeSurvey's time measurements were inaccurate leading us to reconstruct these using the screen captures. 
Two recordings of the semi-structured interviews were corrupted and could therefore not be used in the analysis presented in~\ref{sec:ReasonsForAlgoSelfImpl},
however we were able to refer to our interviewer's notes to code their answers.

Additionally, prior to starting the study on one day, a network issue required the use of a backup laptop with the same software and hardware.                                                      
Therefore, we do not consider this incident to be a data-influencing problem.
No other deviations occurred during the studies execution and all participants completed the experiment.

\newlength{\aboveViolinPlot}
\setlength{\aboveViolinPlot}{-4mm}
\newlength{\belowViolinPlot}
\setlength{\belowViolinPlot}{0mm}
\section{Results and Discussion}
In the following we present the quantitative analysis regarding the influence of the algorithm label on program comprehension and time.
\subsection{\ref{rq:1}: Program Comprehension}
\begin{figure}[h]
	\centering
	\includegraphics[width=\linewidth]{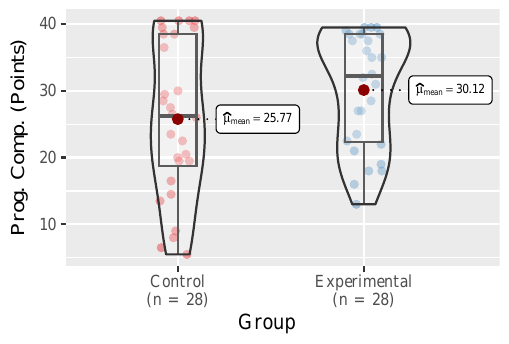}
  \caption{\var{Program comprehension} scores by \var{group}.}
	\label{fig:violin_Plot_Prog_Comp}
  \vspace{\belowViolinPlot}
\end{figure}

In this section, we investigate the impact of algorithm labels on participants' \var{program comprehension}, comparing the performance of the control and experimental groups.
\Cref{tab:desc_stats_Prog_Comprehension} reports the relevant measures regarding the central tendency and dispersion for the \var{program comprehension} of our control and experimental group.
These are also illustrated by the violin plots in \cref{fig:violin_Plot_Prog_Comp}.
\input{r_generated/RQ1ProgramComprehension/desc_statistics_Program_Comprehension.tex}
\vspace{\aboveViolinPlot} 

\subsubsection{Hypothesis Testing}
We perform PERMANCOVA using \var{group} as the independent variable, \var{program comprehension} measured by the achieved points as the dependent variable, and \var{programming experience} based on the pre-questionnaire as the covariate. 
Following Anderson~\cite[p. 74]{Anderson2008} and Bakker~\cite[p. 192-194]{Bakker2024}, we fitted the covariate first and applied Type I sums of squares.
\Cref{tab:permancova_prog_comp} lists the complete output of the model.
\input{r_generated/RQ1ProgramComprehension/permancova_table_program_comprehension.tex}

Looking at the \var{Group} row in \cref{tab:permancova_prog_comp} we can see a significant influence on \var{program comprehension} by the treatment (PERMANCOVA; \Rsq = 0.027, Pseudo-F\textsubscript{1,53} = 3.422, \textit{p}-value = 0.040, perms = 99999).

However, this PERMANCOVA result only tells us that there are differences between the groups, without indicating whether the dissimilarity of the groups stems from differences in the means, differences in the variances or both~\cite[pp. 166-175]{Bakker2024}.
Since the PERMANCOVA results were significant, we also conducted a Permutational Multivariate Analysis of Dispersion 
to assess the homogeneity of variances between the groups (PERMDISP; F\textsubscript{1,53} = 2.53, \textit{p}-value = 0.118).

\begin{figure}[h]
	\centering
	\includegraphics[width=\linewidth]{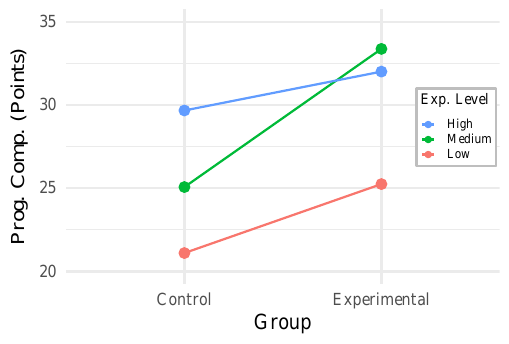}
	\caption{Interaction plot showing the effect of algorithm labels on the mean \var{program comprehension} scores across different levels of prior \var{programming experience}.}
	\label{fig:interaction_plot_Prog_Comp}
\end{figure}

We were also interested in investigating whether participants with different levels of \var{programming experience} responded differently to the algorithm labels.
To explore this, we conducted a post hoc analysis in which we subdivided participants into three equally sized groups (Low, Medium, High) based on their \var{programming experience} scores from the pre-questionnaire. 
This allowed us to create the interaction plot visible in \cref{fig:interaction_plot_Prog_Comp}.

\subsubsection{Interpretation}
  As we can see in \cref{tab:desc_stats_Prog_Comprehension} the experimental group has both a higher mean and a higher median with an improvement by 4.35 and 6.0 points compared to the control group.
  This is also visible in the violin plot which shows that the experimental group has fewer participants that achieved low scores.
  In fact none of the participants in the experimental group scored less than 13 points compared to 5.5 for the control group.
  Additionally, the experimental group has a larger amount of participants that achieved many points.
  This indicates that the algorithm labels help participants that would otherwise achieve low to medium scores to improve their comprehension.
  
  Regarding the dispersion in the groups we can also see that the experimental group has a lower standard deviation.
  This is most likely due to the participants improved points which moves them closer to the center of the distribution.
  However, this difference is not deemed significant by PERMDISP. 

  \Cref{tab:permancova_prog_comp} displays the PERMANCOVA results.
  Analogous to \( \eta^2 \) in traditional MAN(C)OVA, the proportion of variance explained by each factor in the model is reported as \Rsq~\cite[pp. 153-165]{Bakker2024}. 
  We can therefore refer to the guidelines of Cohen~\cite[pp. 284-288]{Cohen1988} for interpreting the effect size~\cite[pp. 491-492]{Field2012},~\cite[p. 33]{dattalo2013analysis}.
  \Cref{tab:permancova_prog_comp} shows that prior \var{programming experience} has a significant influence on the results of the participants with a large effect size according to Cohen~\cite[pp. 284-288]{Cohen1988}.
  This result is unsurprising as  our exercises are based on mental simulation and debugging which are typical activities during programming.
  Participants can therefore leverage their experience to better solve the exercises.
  Additionally, individuals with more programming experience are more likely to recognize the algorithms featured in the exercises, even without the labels.
  This also shows the importance of controlling for prior programming experience to prevent confounding in the results.

  With regard to the treatment we can see that PERMANCOVA also considers the provided algorithm labels to have significant influence, with PERMDISP confirming that this is due to a difference in the mean and not due to differences in dispersion.

  Although the effect size is considered small by Cohen's guidelines~\cite[pp. 284-288]{Cohen1988}, we observe a median improvement of 6 points in the experimental group.
  This corresponds to a \textasciitilde23\% improvement in \var{program comprehension}, which is a notable improvement.
  We therefore reject the null hypothesis and conclude that the correctness of the answers to the given exercises \textit{does significantly differ} if the source code is annotated with labels of the implemented algorithms.
  More specifically the algorithm labels significantly improve participant's program comprehension, reflected in a median score increase of 6 points.

  The interaction diagram in \cref{fig:interaction_plot_Prog_Comp} allows us to compare if the influence of the algorithm labels changes based on the participants prior programming knowledge.
  Firstly, we can see that all experience subgroups benefit from the algorithm labels with their respective group means improving when labels are provided.
  However, the plot also indicates that the impact of the algorithm labels varies depending on the participants' prior \var{programming experience}.
  
  We can see that participants with medium \var{programming experience} have the highest relative improvement by 8.3 points.
  With the improvement through the algorithm labels they are even able to achieve more points than participants with high \var{programming experience} who were not presented with the labels.
  The second-highest improvement occurs among the subgroup with low \var{programming experience}, which showed a mean improvement of 4.1 points.

  One possible explanation could be that these participants were, to some extent, overwhelmed by the exercises, which limited their ability to effectively utilize the algorithm labels. 
  While the labels can aid specific aspects of the comprehension process, such as understanding the algorithm's purpose,
  solving the exercises also requires addressing additional challenges like debugging and implementing the solutions.
  These tasks only benefit indirectly from the improved understanding of the participants and are of course highly dependent on the participants abilities and knowledge.
  Other studies also suggest that overwhelmed participants may show diminished or no effects~\cite{Kather2022, Chen2017}.

  The improvement is less pronounced for participants with high \var{programming experience} with a difference of 2.3 points.
  This is in line with our expectation that participants with high programming experience are more likely to recognize the algorithms and purpose of the code, even without the labels.

  \AnswerBox{1}{
		Annotating source code with algorithm labels significantly improves program comprehension (\textit{p}=0.040), as demonstrated by a median improvement of 6 points (\textasciitilde23\%) in the experimental group. 
    A positive effect on the mean scores is observable across all programming experience levels, with the highest relative improvement seen in participants with medium experience (8.3 points), 
    followed by those with low experience (4.1 points), 
    and smaller gains among highly experienced participants (2.3 points).
	}

\subsection{\ref{rq:2}: Time}
\begin{figure}[h]
	\centering
	\includegraphics[width=\linewidth]{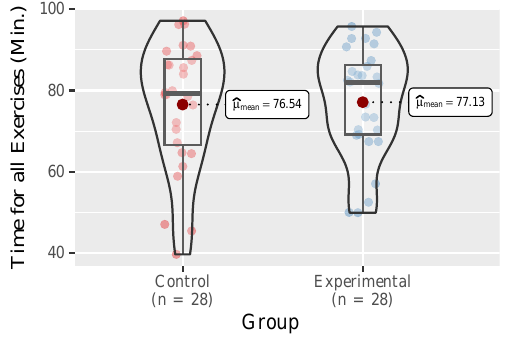}
  \caption{Completion \var{time} across all exercises by \var{group}.}
	\label{fig:violin_Plot_Time}
\end{figure}

Our research question~\ref{rq:2} focuses on whether annotating the source code with algorithm labels reduces the \var{time} required by the participants to solve the exercises.
The measures of central tendency and dispersion for both groups are summarized in \cref{tab:desc_stats_Time}.
A visual representation of these measures is shown in \cref{fig:violin_Plot_Time}.
\input{r_generated/RQ2Time/desc_statistics_Time.tex}
\vspace{\aboveViolinPlot}

\subsubsection{Hypothesis Testing}
We conducted a PERMANCOVA using \var{group} affiliation as independent variable, the \var{time} required by the participants to complete all exercises as dependent variable and \var{programming experience} based on the pre-questionnaire as covariate.
As previously we fitted the covariate first and applied Type I sums of squares, following the recommendations of Anderson~\cite[p. 74]{Anderson2008} and Bakker~\cite[p. 192-194]{Bakker2024}.
\Cref{tab:permancova_time} lists the complete output of the model.
From the \var{Group} row, we observe that PERMANCOVA does not identify a significant effect of the algorithm labels on the completion \var{time}, (PERMANCOVA; \Rsq = 9e-05, Pseudo-F\textsubscript{1,53} = 0.009, \textit{p}-value = 0.991, perms = 99999).
\input{r_generated/RQ2Time/permancova_table_time.tex}

\subsubsection{Interpretation}
Consulting the descriptive statistics in \cref{tab:desc_stats_Time} and the violin plots in \cref{fig:violin_Plot_Time} we observe that both the differences in the mean and median are negligible when compared to the standard deviation.
The violin plots also show quite similar shapes and IQR's for both groups.
Overall half of the participants completed the exercises within 69 to 87 minutes, with a notable subset taking longer, as evidenced by the broader upper sections of the violin plots.

Considering the PERMANCOVA results depicted in \cref{tab:permancova_time} we find that \var{programming experience} has a significant influence on the \var{time} with an effect size considered large by Cohen~\cite[pp. 284-288]{Cohen1988}.
This aligns with our expectations since participants with more programming experience should be faster in solving typical programming exercises.
Moreover, participants with more programming experience are more likely to recognize the algorithms featured in the exercises, even without the labels allowing them to comprehend the code more quickly.

Regarding the treatment the PERMANCOVA results show \textit{no significant effect} of the provided algorithm labels on the overall completion \var{time}.
Consequently, we can \textit{not} reject the null hypothesis concluding that the time required to complete the given exercises \textit{does not significantly differ} if the source code is annotated with labels of the implemented algorithms.
This goes against our initial expectations formulated in \cref{sec:Hypotheses}.
It is also surprising since \var{time} is negatively correlated with \var{program comprehension} (Kendall's \( \tau \) = -0.353, \textit{p}<0.01), meaning that individuals with more points also tend to be faster.
Since the algorithm labels improve the achieved points we would also expect a decrease in the \var{time} required.

\begin{figure*}[h]
	\centering
	\includegraphics[width=0.8\textwidth]{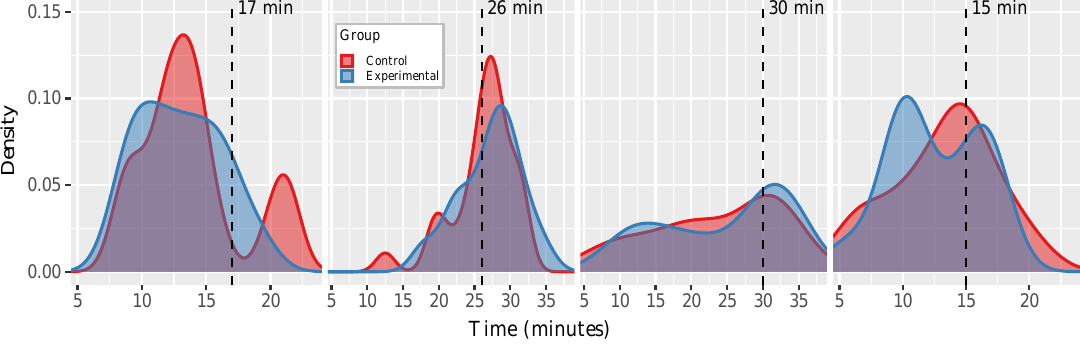} 
  \caption{Distribution of the completion \var{times} for exercise one (leftmost) to exercise four (rightmost). Dashed lines indicate the time limit reminders. Note that the x-axis is scaled differently because some exercises were shorter than others.}
	\label{fig:density_time}
\end{figure*}

We theorized that one possible explanation for this result might the behavior of the participants.
If the majority of participants used the full time allotted for the exercises in an effort to maximize their points, their completion times might not differ significantly, as they would be influenced by the time limits set for each exercise.
Note, that participants were not instructed to work as fast as they can.
Instead, we would ask them to finish the current exercise by writing down their most recent insights and move on to the next exercise once the time limit for the current exercise was reached.
To investigate this further and to analyze potential differences between exercises, we created the density plots shown in \cref{fig:density_time}.
These illustrate the distribution of participants completion \var{times} for each exercise.

Except for the first exercise, which also served as a warm-up, all density plots show a peak at, or shortly after, our time-limit reminder.
This indicates that a notable portion of participants utilized the available time completely and only finished the exercises after or due to the reminder, supporting our theory.
From this we also conclude that the difficulty of the exercises was appropriate and not too easy.

Another interesting insight when examining the plots for Ex3 and Ex4 is the bimodal shape of the distribution observed in the experimental group.
Further analysis revealed that most participants in the experimental group who finished at or before the first peak achieved full marks for the respective exercise.
This further supports our theory that only participants who fully completed the exercises finished early. 
Additionally, this indicates that at least some participants finished earlier due to the provided algorithm labels.

\AnswerBox{2}{
  No statistically significant difference in completion times has been found between groups when annotating the source code with algorithm labels (\textit{p}=0.991).
  This result might have been influenced by the participants' tendency to utilize the entire available time.
  Further analysis of individual exercises supports the notion that the algorithm labels benefited at least some participants in specific exercises.
}

\section{\ref{rq:3} Qualitative Insights} 
In this section we delve into the analysis of the qualitative data collected during our study.
We begin by exploring participants' views on the helpfulness of algorithm labels, along with the explanations and rationale they provide.
We then assess additional use cases for algorithm recognition based on participants' responses, categorizing them to understand the contexts in which they expect to benefit from algorithm recognition.
Finally, we examine under which circumstances algorithms are self-implemented in real codebases rather than using a library.
To analyze the data two researchers qualitatively coded the relevant parts of the post-questionnaire and semi-structured interviews performing \textit{summarizing content analysis} based on the process outlined by Kuckartz~\cite[pp. 58-63]{Kuckartz2014}.

\subsection{Perceived Helpfulness of Algorithm Labels}
\begin{figure}[h]
	\centering
  \includegraphics[width=0.95\linewidth]{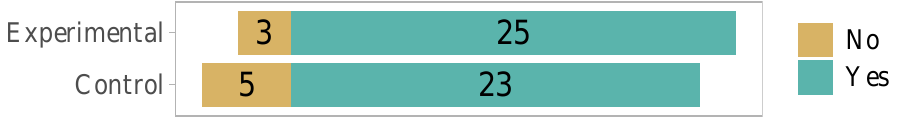}
  \caption{Responses on the helpfulness of algorithm labels.} 
	\label{fig:likert_plot_helpfulness_absolute}
\end{figure}

Our post-questionnaire contained a question regarding the usefulness of algorithm labels, with the question's phrasing depending on the group membership:
\textit{\say{Do you think that labeling the source code with labels of the implemented algorithms [helped/could have helped] you to answer the questions and solve the tasks?}}.
Additionally, we provided the algorithm label for Quicksort as illustration and \dash specifically for the control group \dash explained that these would have been provided for each exercise.
Participants also had the option to provide an explanation for their response  in a separate text field.
Note that the goal of the study was only communicated to the participants at this point.
Prior to this, we had advertised it as a program comprehension study to prevent participants from giving the labels excessive attention.
\Cref{fig:likert_plot_helpfulness_absolute} shows the participants answers per group.

As we can see most of the participants in the experimental group found the labels helpful.
Of the three participants in the experimental group who did not find the labels helpful, 
two stated they could identify the algorithms in the exercises without them but noted the labels would help with unfamiliar ones.
The third participant chose \textit{\say{No}} because they did not notice the algorithm labels.
The overall picture regarding the opinions of the control group is quite similar.
Of the five participants that answered with \textit{\say{No}} four conveyed that recognizing the algorithms was not the main issue for them.

\Cref{fig:code_book_helpfulness} gives an overview of \textit{why} participants did (not) find the labels helpful, with the numbers denoting how often a category was mentioned by different participants.
Note that participants could give multiple reasons, including both positive and negative.
\newcommand{\codeBookZoom}{0.88}
\begin{figure}[h]
  \vspace*{-2mm}
	\centering
  \includegraphics[width=\codeBookZoom\linewidth]{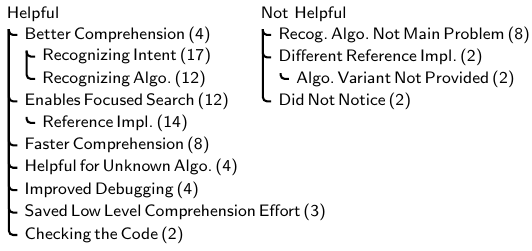}
  \caption{Coding categories for the helpfulness of labels.}
	\label{fig:code_book_helpfulness}
\end{figure} 

Regarding the reasons why participants did not benefit from the labels, eight answered that recognizing the algorithms was not the main problem for them.
Some participants were also unsure or confused by the different implementation variants presented online with two noting that including the specific implementation type in the labels would enhance their usefulness.
Two participants mentioned that they did not notice the labels.

With respect to the positive impact of the algorithm labels the participants expressed that the labels enabled them to better comprehend the code through understanding both the overall purpose and intent of the code as well as the individual algorithms.
To quote one of the participants: \textit{\say{The code comprehension is greatly improved if the underlying idea behind the code is known beforehand.}}\footnote{Where necessary direct quotes have been literally translated from German into English.}. 
Other responses also expressed that the labels enabled them to search for additional information online to better understand the algorithms, with many mentioning that reference implementations or pseudocode were particularly useful.
Participants also noted that they were able to faster comprehend the code and that the labels were especially helpful for unknown algorithms.
Other responses indicated that the labels improved debugging through a better understanding of \textit{\say{what [the code] is supposed to do in each step.}}.
Finally, participants also mentioned that the labels enabled them to avoid low-level comprehension effort when inferring the meaning of the code, and allowed them to run the code along a mental checklist to verify whether it implemented the algorithm annotated by the label.
This aligns with the concept of top-down comprehension~\cite{Shaft1995, Siegmund2016} where developers recognize familiar patterns, quickly validate them, and infer the code's meaning instead of resorting to time-consuming bottom-up comprehension.

Some participants suggested additional improvements to enhance the utility of the labels.  
These included specifying the type of implementation directly in the label, adding comments to explain individual algorithm steps and the purpose of variables as well as providing a brief general explanation of the algorithm.
Providing additional details, such as time and memory complexity or stability for sorting algorithms, were also mentioned.
While we deliberately kept the labels minimal to assess whether they would be beneficial on their own, future research could focus on the influence of these improvements on program comprehension.

\subsection{Use Cases for Algorithm Recognition}
\begin{figure}[h]
	\centering
  \includegraphics[width=\codeBookZoom\linewidth]{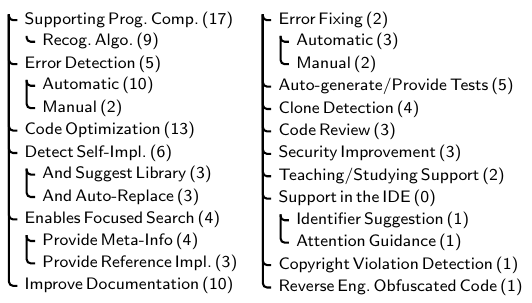}
  \caption{Coding categories of algorithm recognition use cases.}
	\label{fig:code_book_use_cases} 
\end{figure}

We were also interested to investigate in which broader applications' algorithm recognition could be helpful. 
We therefore asked the participants \textit{\say{In which use cases do you think automatic algorithm recognition could be helpful?}}. 
\Cref{fig:code_book_use_cases} gives an overview of the responses, with the numbers denoting how often a category was mentioned by different participants.

Unsurprisingly some of the use cases mentioned overlap with the reasons participants gave for the helpfulness of algorithm labels such as using them to support program comprehension in general or to enable focused search.
We therefore focus on the other categories in the following.
In total fifteen participants mentioned error detection with ten envisioning it as an automatic process while two specified a tool-supported process such as diffing of recognized algorithms with correct implementations to spot bugs.
Five responses did not provide details on this distinction.

Other participants focused on the possibility of optimizing existing self-implementations by suggesting either common optimizations of algorithmic procedures or entirely different algorithms that would be better suited for the task.
Further interest was expressed in detecting self-implementations, often with the goal of (automatically) replacing them with library functions.
Participants also expressed interest in improving or auto-generating documentation as well as automated or tool-supported error fixing.
With respect to code quality, generating or providing test cases were also suggested to ensure the correctness of the recognized algorithm implementations.

Supporting clone detection on a higher level of abstraction as well as facilitating code reviews were also seen beneficial use cases.
Participants also noted the possibility of improving the security of code by recognizing unsafe practices such as (wrongly) self-implemented input validation or use of (unsecure) legacy functions.
Leveraging algorithm recognition in IDEs to propose better identifiers or guide developers' focus by highlighting or de-emphasizing code sections were also suggested.
Lastly, participants mentioned detecting copyright violations and aiding in the reverse engineering of obfuscated code.

In conclusion, participants shared many interesting use cases in which they expected to benefit from algorithm recognition. 
Aside from library replacement (as explored, for example, by Metzger et al.~\cite{Metzger2000}) and teaching support (as discussed by Nunez et al. and Taherkhani et al.),
few of these use cases have been addressed by algorithm recognition tools.
These insights provide valuable future avenues for the development of tools based on algorithm recognition as well as future experimental studies regarding their usefulness.

\subsection{Reasons for Algorithm Self-Implementation}\label{sec:ReasonsForAlgoSelfImpl} 
To better understand self-implementation practices, we explored which algorithms were self-implemented in practice and the motivations behind this.

In total, 15 participants reported that algorithms were self\hyph implemented in code bases they had insight into, with one noting that these implementations were later replaced by library code.
Four participants reported no self-implementations of algorithms in the codebases they knew, citing reasons such as the applications being purely frontend or data-focused (i.e. aimed at transforming and visualizing data).
Three participants did not have insights into real-world code bases.
The topic was not discussed in the remaining interviews.

When discussing reasons for self-implementations six participants cited missing library support with some specifying legacy code such as Cobol being problematic in this respect. 
Library inadequacies were cited five times, including issues such as incompatible data types, special edge cases that could not be handled by the library, problems when processing large amounts of data and discontinued library maintenance.
Participants also explained actively avoiding library dependencies due to licensing costs or the required code being too simple to justify including all the dependencies of a library.
Finally, performance requirements, better maintainability as well as the special case of a re-implementation of a library interface that provided certain algorithms were also mentioned.

\begin{figure}[h]
	\centering
  \includegraphics[width=\codeBookZoom\linewidth]{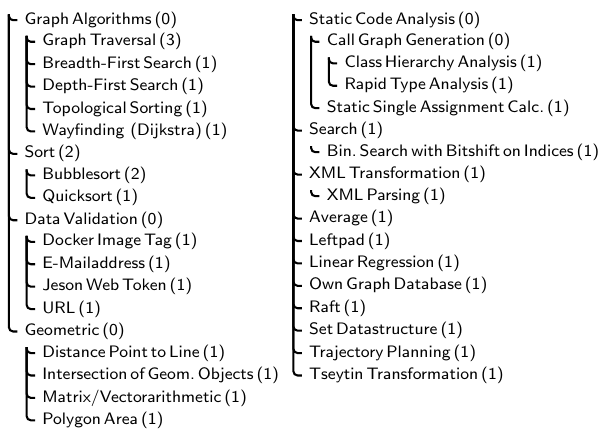}
  \caption{Coding categories for the participant-reported examples of self-implemented algorithms in their codebases.}
	\label{fig:code_book_examples_of_algorithms} 
\end{figure}

The examples brought up by the participants are displayed in \cref{fig:code_book_examples_of_algorithms}.
In total participants mentioned 35 examples from a wide variety of application areas.
These included classic use cases such as sorting, searching, graph algorithms, and geometric calculations, as well as specialized tasks like data validation and static analysis.
Interestingly the specific algorithms mentioned also covered a large spectrum ranging from basic algorithms like Bubblesort, over more complicated matrix-multiplications to specific and complex algorithms like Raft for consensus in distributed systems.

In conclusion, algorithm self-implementation is surprisingly common, driven by reasons such as missing library support, deliberate avoidance of library dependencies, and specific requirements like performance.
The implemented algorithms span multiple domains (e.g. graph handling, geometric calculations) and range from simple (Bubblesort) to complex (Raft).
These findings suggest that algorithm recognition can be helpful in real-world codebases.
While tools can and should prioritize recognizing algorithms in the identified domains, further research is needed to better quantify which types of algorithms and domains are most relevant in practice.

\AnswerBox{3}{
Most participants found the algorithm labels helpful, as they enabled information retrieval and enhanced comprehension of both the algorithms and the overall purpose of the code.
Participants also outlined many potential use cases for algorithm recognition including error detection, code optimization and library replacement.
Self-implemented algorithms are surprisingly common, often due to missing library support, dependency concerns, or performance requirements.
Examples of self-implementations range from simple algorithms like Bubblesort to complex ones such as Raft.
}

\section{Threats to Validity}
    To ensure \textbf{construct validity} we designed our study following the guidelines of Siegmund~\cite{Siegmund2012a} and Dunsmore et al.~\cite{Dunsmore2000} operationalizing \var{program comprehension} via the correctness of mental simulation and maintenance exercises. 
    We also designed the scoring such that the achievable points scale with question difficulty, ensuring the ratio-scale properties of our measurements.

    Concerning \textbf{internal validity}, we controlled for \var{programming experience} as a confounder, including it in our statistical analysis. 
    We also considered other potential confounders, such as Java and Eclipse experience, with the former being comparably distributed among groups and the latter not correlating with the dependent variables. 
    Additionally, two pilot studies helped to identify potential issues in the study design, question formulation, and execution.

    Regarding the standardization of study setup, we developed a \say{pre-flight checklist} ensuring uniform conditions across sessions and participants.
    To promote adherence to guidelines, at least one researcher was always present during the sessions,
    and both experimenters conducted the initial sessions together to establish a shared understanding of the procedures. 

    For an objective evaluation, a sample solution and written assessment instructions were developed.
    Two researchers independently graded the exercises and performed qualitative coding,
    resolving discrepancies through discussion and updating guidelines where needed.
    To evaluate the maintenance exercises the test suite was executed, and the solutions were additionally reviewed manually.

    Regarding \textbf{conclusion validity} we employed robust statistical tests, carefully checked their assumptions, and adhered to best practices outlined by Anderson~\cite{Anderson2008} and Bakker~\cite{Bakker2024}.
    Additionally, the quantitative results also align with participants' responses, with a large majority reporting that the labels were helpful.
    
    \Cref{fig:interaction_plot_Prog_Comp} divides participants into smaller experience groups (\textasciitilde 9 participants each), which is a threat.
    However, we see a clear ordering from Low to High, consistent with expectations.
    We also limit our analysis to the insight that all experience groups benefit from the labels, with the medium group benefiting most.

    The time-related finding should be interpreted cautiously, as participants may have intentionally used the full available time to maximize their points,
    with at least some of the participants' reporting that they comprehended the code more quickly thanks to the labels.

    One threat to the \textbf{external validity} of our results is the fact that approximately half of our participants are students.
    However, Falessi et al. argue that using students effectively simulates real software engineering settings in laboratory contexts~\cite{Falessi2018}. 
    Moreover, our study also included researchers with more work and programming experience than students, as well as professional software developers.

    The source code base used in our experiments is implemented in Java which is another limitation.
    However, we argue that Java is one of the most widely used programming languages and a good representative for object-oriented programming languages.
    Furthermore, Java was the language our participants were most experienced with, enhancing the validity of our experiments.
    The code base was also specifically designed to facilitate realistic experiments, 
    with our exercises reflecting practical functionalities within the system and being representative of typical programming tasks.

    Finally, our exercises included algorithms of easy to moderate complexity, that are often covered in curricula and are thus among the better-known ones.
    However, since additional information aids their comprehension, we also expect it to be beneficial for more complex algorithms as they are harder to understand by definition.

\section{Related Work} 
To establish context, we review work on program concept recognition followed by an examination of program comprehension.

\textbf{Program Concept Recognition:}
Various approaches aim to recognize concepts in code~\cite{Kozaczynski1992, Quilici1994, Wills1994, Nunez2017, ARCC2}, including coding strategies, data structures, and architectural patterns, with some specifically targeting algorithm recognition~\cite{Metzger2000, Alias2003, Zhu2011, Taherkhani2013, Mesnard2016, Nunez2017, ARCC2, Long2022, Neumueller2024}.
Common methods for addressing these challenges involve graph matching of predefined search patterns within code representations (e.g., abstract syntax trees or data-flow graphs)
and employing machine learning classifiers with extracted code features or vectorized form of the aforementioned code representations.
However, to the best of our knowledge, these approaches are not evaluated with respect to their impact on program comprehension.
With the few existing evaluations only addressing technical aspects like scalability or performance.
This gap, also emphasized by Storey et al.~\cite{Storey2000} who highlight the lack of empirical studies on comprehension tools and features that enhance program understanding, 
along with no evaluation of existing approaches, motivates our study.

\textbf{Program Comprehension:}
Numerous studies have explored how programmers comprehend source code~\cite{Shneiderman1979, Pennington1987, Shaft1995, Mayrhauser1995, Siegmund2016, Wyrich2023}.
The prevailing consensus is that developers combine both top-down and bottom-up strategies~\cite{Shaft1995, Siegmund2016}.
Programmers prefer the top-down strategy, recognizing familiar concepts in the code to form and verify hypotheses about the program's purpose.
When programmers fail to recognize familiar concepts, they adopt a bottom-up strategy reading the code line by line and slowly summarizing them into higher level abstractions~\cite{Shaft1995, Siegmund2016}.
Subsequent research also demonstrates that programs utilizing typical patterns are easier to understand~\cite{Soloway1984, Siegmund2016}.
Algorithms can be viewed as specific instances of these patterns.
Indeed, among the various responses for the helpfulness of labels, several participants indicated that the labels helped them recognize the algorithms and apply a top-down comprehension technique.

Kather et al.~\cite{Kather2021} investigate the connection between program and algorithm comprehension to improving algorithm teaching.
Their qualitative study with (PhD) students examines how algorithms, presented as pseudocode with textbook explanations, are understood, 
revealing similarities in comprehension techniques but also key differences, such as the need to understand proofs.
Their finding that even experienced participants rely on additional prose for comprehension, aligns with our hypothesis that extra information enhances understanding.
However, our study shifts the focus to \textit{program comprehension}, using both qualitative and quantitative methods and includes industry developers working on realistic coding tasks.

To summarize, there is a substantial body of literature on program comprehension, which focuses on various aspects of code comprehension.
Nevertheless, to the best of our knowledge, none of the aforementioned papers evaluate if providing additional information on the implemented algorithms improves program comprehension.

\section{Conclusions and Future Work}
This paper presents a controlled experiment to evaluate whether annotating source code with algorithm labels improves program comprehension in terms of correctness and time.
We compared two groups, one receiving algorithm labels as support, and measured their performance through program comprehension exercises.
The results show that algorithm labels improved \var{program comprehension}, with a 6-point (23\%) median score increase, particularly benefiting participants with medium and low \var{programming experience}.
No significant difference in the completion \var{times} between the groups was observed, likely because participants tended to use the entire available time to maximize their scores.
Interestingly exercise-level analysis as well as the participant responses indicate that at least some participants benefited regarding the completion \var{time}.
Qualitative analysis of the responses finds that a majority of participants perceived the labels as helpful. 
Other use cases in which participants expected to benefit from algorithm recognition included error detection, code optimization or the generation of test cases.
Lastly participants shared reasons for self-implementing algorithms which included library inadequacies, performance requirements, dependencies and licensing costs. 

More research is needed to fully assess the influence of algorithm labels on the time required to comprehend programs.
Additionally, the insights from our study, such as the use cases mentioned by participants, provide valuable avenues for future research, 
not only for tool builders but also for the broader exploration of how algorithm recognition can enhance various aspects of software development.

\ifanonymous
\else
  \section{Acknowledgments}
   We would like to thank Fenja Maier for her contributions to the design and execution of the user study.
   We also thank all participants for taking part in the study.
\fi

\bibliographystyle{ACM-Reference-Format}
\bibliography{references}

\end{document}

%% file: r_generated/RQ1ProgramComprehension/desc_statistics_Program_Comprehension.tex
\setlength{\tabcolsep}{4pt} 

\def\head#1{\multicolumn{1}{c}{\kern-3pt#1}} 
\begin{table}[hb]
\centering
\begin{tabular}{lr*6{@{\hskip5pt}S}} 
  \hline
Group & \head{n} & \head{Mean} & \head{Median} & \head{SD} & \head{Min} & \head{Max} & \head{Skew} \\ 
  \hline
Control &  28 & 25.77 & 26.25 & 11.59 & 5.50 & 40.50 & -0.17 \\ 
  Experimental &  28 & 30.12 & 32.25 & 8.62 & 13.00 & 39.50 & -0.45 \\ 
   \hline
\end{tabular}
\caption{Statistics for the variable \var{program comprehension.}} 
\label{tab:desc_stats_Prog_Comprehension}
\end{table}

\setlength{\tabcolsep}{6pt}

%% file: r_generated/RQ1ProgramComprehension/permancova_table_program_comprehension.tex
\begin{table}[hb]
  \centering
  \begin{tabular}{*5{r@{\hskip10pt}}r}
    \hline
   & \vphantom{\large I}Df & SumOfSqs & \RsqTable & F & \textit{p}-value \\ 
    \hline
  Prog. Experience & 1 & 61.373 & 0.558 & 71.212 & 1e-05 \\ 
    Group & 1 & 2.949 & 0.027 & 3.422 & 0.040 \\ 
    Residual & 53 & 45.678 & 0.415 &  &  \\ 
    Total & 55 & 110.000 & 1.000 &  &  \\ 
     \hline
  \end{tabular}
  \caption{PERMANCOVA results for \var{program comprehension}, showing the effects of \var{programming experience} and \var{group}.} 
  \label{tab:permancova_prog_comp}
  \end{table}

%% file: r_generated/RQ2Time/desc_statistics_Time.tex
\setlength{\tabcolsep}{4pt} 

\def\head#1{\multicolumn{1}{c}{\kern-3pt#1}} 
\begin{table}[ht]
\centering
\begin{tabular}{lr*6{@{\hskip5pt}S}}
  \hline
  Group & \head{n} & \head{Mean} & \head{Median} & \head{SD} & \head{Min} & \head{Max} & \head{Skew} \\ 
  \hline
Control &  28 & 76.54 & 79.24 & 15.62 & 39.69 & 97.13 & -0.76 \\ 
  Experimental &  28 & 77.13 & 81.91 & 13.45 & 49.94 & 95.74 & -0.57 \\ 
   \hline
\end{tabular}
\caption{Statistics for the variable \var{time} in minutes.} 
\label{tab:desc_stats_Time}
\end{table}

  \setlength{\tabcolsep}{6pt}

%% file: r_generated/RQ2Time/permancova_table_time.tex
\begin{table}[ht]
\centering
\begin{tabular}{*5{r@{\hskip10pt}}r}
  \hline
  & \vphantom{\large I}Df & SumOfSqs & \RsqTable & F & \textit{p}-value \\ 
  \hline
Prog. Experience & 1 & 55.984 & 0.509 & 54.940 & 1e-05 \\ 
  Group & 1 & 0.009 & 9e-05 & 0.009 & 0.991 \\ 
  Residual & 53 & 54.007 & 0.491 &  &  \\ 
  Total & 55 & 110.000 & 1.000 &  &  \\ 
   \hline
\end{tabular}
\caption{PERMANCOVA results for \var{time}, showing the effects of \var{programming experience} and \var{group} affiliation.} 
\label{tab:permancova_time}
\end{table}